\pdfoutput=1
\RequirePackage{ifpdf}
\ifpdf 
\documentclass[pdftex]{sigma}
\else
\documentclass{sigma}
\fi

\begin{document}

\newcommand{\arXivNumber}{1507.02201}

\allowdisplaybreaks

\renewcommand{\PaperNumber}{071}

\FirstPageHeading

\ShortArticleName{Path Integrals on Euclidean Space Forms}

\ArticleName{Path Integrals on Euclidean Space Forms}

\Author{Guillermo CAPOBIANCO and Walter REARTES}
\AuthorNameForHeading{G.~Capobianco and W.~Reartes}

\Address{Departamento de Matem\'atica, Universidad Nacional del Sur,\\
 Av.~Alem 1253, 8000 Bah\'ia Blanca, Buenos Aires, Argentina}
\Email{\href{mailto:guillermo.capobianco@gmail.com}{guillermo.capobianco@gmail.com}, \href{mailto:walter.reartes@gmail.com}{walter.reartes@gmail.com}}

\ArticleDates{Received July 09, 2015, in f\/inal form August 31, 2015; Published online September 03, 2015}

\Abstract{In this paper we develop a quantization method for f\/lat compact manifolds based on path integrals. In this method the Hilbert space of holomorphic functions in the complexif\/ication of the manifold is used. This space is a reproducing kernel Hilbert space. A def\/inition of the Feynman propagator, based on the reproducing property of this space, is proposed. In the $\mathbb{R}^n$ case the obtained results coincide with the known expressions.}


\Classification{53Z05; 81S40}

\section{Introduction}

The quantization of a system whose conf\/iguration space is a dif\/ferentiable manifold is a far from exhausted problem. For instance, in the case of a Riemannian manifold, it is not known which quantization scheme best represents the curvature of the manifold, see \cite{casimir31, chaichian01, dewitt57, kleinert89, kleinert90, kleinert04,kowalski-rembielinski00, kowalski-et-al98, mostafazadeh96}. Such problems have both physical and mathematical motivations. On one side there is the problem of the existence of the structures involved in the quantization and, on the other, the possible applications to specif\/ic physical problems.

When the manifold is a compact Lie group, Hall showed that the quantization of the group is naturally isomorphic to the quantization of the cotangent space of the group, i.e., a quantization on the phase space. The latter coincides with the complexif\/ication of the group, see \cite{hall94, hall00b, hall05}. In other words, there are two Hilbert space structures, one given by the functions on the group and the other by the functions in the complexif\/ication of the group, both structures being naturally related. These problems have been explored and generalized in several other works, see, e.g., \cite{hall94, hall97a, hall00b, hall05, stenzel99}.

It is known that the cotangent bundle of a Riemannian manifold admits a natural complex structure, compatible with the symplectic form and the natural lifting of the Riemannian metric if and only if the manifold is f\/lat. In the Appendix~\ref{appendix} we show this result following the work of Gorbunov~\cite{gorbunov-et-al05}, see also~\cite{kobayashi69-2}.

In the case of an orientable connected compact f\/lat Riemannian manifold (Euclidean space form) we show that there is a natural isomorphism between the Hilbert space of square integrable complex functions on the conf\/iguration space and the space of square integrable holomorphic functions on the phase space. The scalar products are def\/ined with a measure given by the fundamental solution of the heat equation on each space.

This space of holomorphic functions on the phase space turns out to be a reproducing kernel Hilbert space. Taking advantage of the existence of a reproducing kernel we obtain the above mentioned isomorphism and a path integral which coincides with the known expressions in the Euclidean case, see \cite{deligne-et-al99, zinnjustin05}.

In particular, the $3$-dimensional orientable compact Euclidean space forms present a~parti\-cu\-lar interest for cosmology, since they could model the spatial part of the f\/lat-universe models~\cite{ellis71}.
See the most recent works of J.~Levin et al., which seek to develop a plausible cosmological model using orientable compact Euclidean space forms of dimension $3$ in agreement with results of observations made on the cosmic  microwave background radiation~\cite{levin02, levin98,levin98_2,levin98_3}.

\section{Flat Riemannian manifolds}

A connected complete f\/lat Riemannian manifold is the quotient of $\mathbb{R}^n$ by a subgroup $\Gamma$ of the Euclidean group~$E(n)$, which has a free and properly discontinuous action. This is a part of a~more general theorem by W.~Killing and H.~Hopf~\cite{wolf06}.

\begin{theorem}
Let $M$ be a Riemannian manifold of dimension $n\geq2$ and zero curvature. Then~$M$ is complete and connected if and only if it is isometric to the quotient $\mathbb{R}^n/\Gamma$ with $\Gamma \subset E(n)$, where~$\Gamma$ acts freely and properly discontinuously.
\end{theorem}

These manifolds are known as Euclidean space forms. The Euclidean group $E(n)$ is the semidirect product of the groups~$O(n)$ and $\mathbb{R}^n$. An element $\gamma\in\Gamma\subset E(n)$ is identif\/ied with $\gamma=(A,a)$, $A\in O(n)$ and $a\in\mathbb{R}^n$. The action of this group on an element $x\in\mathbb{R}^n$ is given by $\gamma(x)=A x+a$.

In one dimension the manifolds of this type are the real line $\mathbb{R}$ and the circle $S^1$. In dimension~$2$ there are f\/ive manifolds, the plane~$\mathbb{R}^2$, the cylinder, the inf\/inite M\"oebius strip, the torus and the Klein bottle. The torus and Klein bottle are both compact, while the torus is the only orientable one. In dimension $3$ there are $18$ types,  $10$ of which are compact, $6$~orientable and $4$ non-orientable~\cite{kuhnel06,wolf06}. In higher dimensions the number grows signif\/icantly; for example, there are $74$ compact types in dimension~$4$.

If $\Gamma \subset E(n)$ is a lattice, then the quotient $\mathbb{R}^n/\Gamma$ is called an $n$-torus, and is a compact Euclidean space form and a Lie group.

It can be shown that every homogeneous Riemannian manifold is dif\/feomorphic to some Lie group but in general a space form is not necessarily homogeneous. In particular, when the space form is homogeneous we have the following theorem~\cite[p.~88]{wolf06}:

\begin{theorem}
Let $M$ be a connected homogeneous Riemannian manifold of dimension~$n$ and zero curvature, then it is isometric to the product $\mathbb{R}^m\times T^{n-m}$ of a Euclidean space with a flat Riemannian torus.
\end{theorem}

A discrete subgroup is a subgroup which is a discrete subset. If $\Gamma$ is a closed subgroup of~$G$, is called uniform if the quotient space $G/\Gamma=\{g\Gamma\colon g \in G\}$ is compact.

The following theorem characterize discontinuous groups on Euclidean spaces.

\begin{theorem}\label{t}
Let $\Gamma$ be a subgroup of the Euclidean group $E(n)$.
\begin{itemize}\itemsep=0pt
\item[$(i)$] $\Gamma$ acts properly discontinuously on Euclidean space~$\mathbb{R}^n$ if, and only if, $\Gamma$ is discrete on~$E(n)$.
\item[$(ii)$] If $\Gamma$ is closed, then it acts freely on~$\mathbb{R}^n$ if and only it is torsion free.
\item[$(iii)$] $\Gamma$ acts properly discontinuously and with compact quotient on~$\mathbb{R}^n$ if, and only if, $\Gamma$ is a~discrete uniform subgroup of~$E(n)$.
\end{itemize}
\end{theorem}

This paper focuses on orientable compact f\/lat manifolds.

Flat compact Riemannian manifolds of dimension $n$ are quotients of polyhedra in~$\mathbb{R}^n$ by identifying faces (see \cite[p.~99]{wolf06}). The interior of these polyhedra may be taken as a chart, which we call~$Q^\circ$. Functions def\/ined on the manifold are functions on~$\mathbb{R}^n$, which are invariant under the action of the group.

An important invariant for a compact Euclidean space form is its volume. This can be def\/ined in terms of a fundamental region for~$\Gamma$ in~$\mathbb{R}^n$~\cite{mcmullen02}.
The volume of a space form $\mathbb{R}^n/\Gamma$ is def\/ined to be the volume of any
fundamental region.
As a fundamental region for $\Gamma$ we can take $c_\gamma$, the closure of the Dirichlet domain centered at~$\gamma (0)$
\begin{gather*}
c_\gamma:=\big\{x\in \mathbb{R}^n; \|\gamma(0)-x\|\leq\|\gamma'(0)-x\|\,\text{for every}\, \gamma' \in \Gamma\big\},
\end{gather*}
where $\gamma (0)$ is the action of $\gamma$ on $0\in\mathbb{R}^n$. $c_\gamma$ is an $n$-dimensional convex polyhedron in $\mathbb{R}^n$ bounded by hyperplanes which are perpendicular bisectors of line segments $[\gamma(0), \gamma'(0)]$. Its bounda\-ry~$\partial c_\gamma$ carries a locally f\/inite decomposition into convex polyhedra of dimension $n-1$. The space form~$\mathbb{R}^n/\Gamma$ is then obtained from~$c_\gamma$ by identifying points in~$\partial c_\gamma$ which are equivalent modulo~$\Gamma$.

In particular, the six $3$-dimensional orientable compact Euclidean space forms are the following quotient spaces $\mathbb{R}^3/\Gamma_i$, $i=1,\dots,6$ (see \cite[p.~117]{wolf06} and \cite[p.~302]{kuhnel06}). The torus $T^3$, which is constructed by identifying the opposite faces of a parallelepiped by translations, in this case $\Gamma_1$ is generated by three translations $t_1$, $t_2$, $t_3$, in the direction of three linear independent vectors. Other four are obtained after gluing with a quarter turn, a half-turn, a one-sixth turn and a one-third turn. The last one is the Hantzsche--Wendt space which has a more complicated structure, $3$ screw motions are needed. $\Gamma_2$ is generated by $\Gamma_1$ and a screw motion $\alpha^2=t_3$, the faces of the translated parallelepiped are identif\/ied after a rotation of an angle of~$\pi$, $\Gamma_3$  is generated by $\Gamma_1$ and a screw motion $\alpha^3=t_3$. $\Gamma_4$ is generated by $\Gamma_1$ and a screw motion~$\alpha^4=t_3$. $\Gamma_5$ is generated by~$\Gamma_1$ and a screw motion $\alpha^6=t_3$. In the last one, the Hantzsche--Wendt space, $\Gamma_6$ results from $\Gamma_2$ by adding two further screw motions by an angle of~$\pi$. The manifolds $\mathbb{R}^3/\Gamma_3$ and $\mathbb{R}^3/\Gamma_5$ are obtained from a lattice made by translating a hexagonal plane lattice a certain distance perpendicular to the plane and identifying opposite sides with the top rotated by $2\pi/3$ and $\pi/3$ respectively.

The family $\{c_\gamma\}$ forms a crystalline structure whose symmetry group contains $\Gamma$ as a subgroup of f\/inite index (see \cite[p.~100]{wolf06}). We can choose the interior of~$c_0$ (the cell corresponding to the identity of the group) as the chart $Q^\circ$ of $\mathbb{R}^n/\Gamma$.
The crystalline structure can be generated by translation of a f\/inite set of vectors def\/ining the crystal lattice. This set forms a basis of~$\mathbb{R}^n$. Dual basis vectors multiplied by $2\pi$ are the basis of the reciprocal lattice,~$\mathcal{L}$. Let~$K$ be an element of the reciprocal lattice, then a function with the symmetry of this lattice has a Fourier expansion given by
\begin{gather*}
f(x)=\sum_{K\in\mathcal{L}} c_K \mathrm{e}^{\mathrm{i}  K\cdot x}.
\end{gather*}
This function is well def\/ined on the manifold if it is also invariant under the action of $\Gamma$, i.e., $(\gamma f)(x)=f(\gamma x)=f(x)$ for all $\gamma\in\Gamma$.

\section{Hilbert spaces of quantization}

In a series of papers, Hall \cite{hall94, hall00b, hall05} showed that in a compact Lie group there is a natural isomorphism between the space of square integrable functions on the group, with a measure given by the fundamental solution of the heat equation at the identity, and the space of square integrable holomorphic functions in the complexif\/ication of the group (which is isomorphic to the cotangent space). Dif\/ferent but equivalent versions of this isomorphism can be found in the literature, see, e.g., \cite{hall94, hall97a, hall00b, hall05, stenzel99}.

For orientable compact Euclidean space forms we develop this isomorphism explicitly through integrals in Euclidean space and the solutions of the heat equations. Doing so paves the way for us to write an expression for the path integral in these spaces, which is the main result of this work.

\subsection{Weighted heat kernel representation}

First we f\/ind the fundamental solution of the heat equation (heat kernel) in a compact f\/lat Riemannian manifold. The solution of the heat equation def\/ined on the manifold can be calculated by f\/inding the solution on the chart given by the polyhedron $Q^\circ$. Those solutions of the heat equation that are invariant under the action of the subgroup $\Gamma$ are solutions on the manifold.

If the manifold is $\mathbb{R}^n$, the heat equation
\begin{gather}\label{ecuaciondelcaloreuclidiana}
\frac{\partial \rho_t^{x_0}(x)}{\partial t}=\frac{1}{2}\Delta \rho_t^{x_0}(x),
\end{gather}
where $\Delta$ is the Laplacian, has the following fundamental solution
\begin{gather}\label{rhor}
\rho_t^{x_0}(x)=\frac{1}{(2\pi t)^{n/2}} \mathrm{e}^{-(x-x_0)^2/2t}.
\end{gather}
This solution verif\/ies the condition
\begin{gather}\label{delta}
\lim_{t\to 0^+}\rho_t^{x_0}(x) = \delta(x-x_0).
\end{gather}

In the orientable compact case, this equation can be solved on the chart~$Q^\circ$, where we choose a~point~$x_0\in Q^\circ$. We seek a function $\rho_t^{x_0}(x)$ that verif\/ies (\ref{ecuaciondelcaloreuclidiana}) with the condition~(\ref{delta}). As was discussed above, the solution must be of the form
\begin{gather}\label{fundamental}
\rho_t^{x_0}(x)=\frac{1}{V}\sum_{K\in\mathcal{L}}c_K(t) \mathrm{e}^{\mathrm{i}  K\cdot (x-x_0)},
\end{gather}
where we introduced the volume $V$ of the cell because of normalization issues. Also, $\rho_t^{x_0}(x)$ must be invariant under the action of~$\Gamma$. Inserting the last expression in equation~(\ref{ecuaciondelcaloreuclidiana}) and considering~(\ref{delta}), we have
\begin{gather}\label{kernelenq}
\rho_t^{x_0}(x)=\frac{1}{V}\sum_{K\in\mathcal{L}}\mathrm{e}^{\mathrm{i} K\cdot (x-x_0)-K^2t/2}.
\end{gather}
This expression is invariant under the group action. It is a consequence of the symmetry of the coef\/f\/icients. Indeed, if $\gamma=(A,a)\in\Gamma$, then
\begin{gather*}
(\gamma\rho_t)^{x_0}(x)=\frac{1}{V}\sum_{K\in\mathcal{L}}\mathrm{e}^{\mathrm{i} K\cdot\gamma (x-x_0)-K^2t/2}
        =\frac{1}{V}\sum_{K\in\mathcal{L}}\mathrm{e}^{\mathrm{i} K\cdot A (x-x_0) + \mathrm{i} K\cdot a-K^2t/2} =\rho_t^{x_0}(x).
\end{gather*}
The last equality results from the orthogonality of~$A$ and $\Gamma$ having f\/inite index in the symmetry group of the crystal. Then the fundamental solution in $Q^\circ$ is the fundamental solution in~$Q$.

We def\/ine a scalar product on $Q$ by the following expression
\begin{gather*}
\langle f,g\rangle_Q = \int_{Q} \overline{f(x)}g(x)\rho_t^{x_0}(x)\mathrm{d} x.
\end{gather*}
This product gives a Hilbert space which we call $L^2(Q,\rho_t^{x_0})$. We note that this product depends on the point~$x_0$, it is centered on a point of the manifold. The usual representation of quantum mechanics involves the space~$L^2(Q)$ where the integration can be performed on the polyhedron~$Q^0$ with the Lebesgue measure. The isometry between the two representations is given by
\begin{gather}\label{iso}
f(x)=\frac{f_S(x)}{\sqrt{\rho_t^{x_0}(x)}},
\end{gather}
where $f_S$ is the corresponding function on~$L^2(Q)$ and~$f$ on~$L^2(Q,\rho_t^{x_0})$. Here, we name the representation on $L^2(Q,\rho_t^{x_0})$ weighted heat kernel representation.

\subsection{Holomorphic representation}

The cotangent bundle of a f\/lat manifold has a natural complex structure. See the Appendix~\ref{appendix}.

We have seen that an appropriate chart for these manifolds is the interior of a polyhedron in~$\mathbb{R}^n$, which we call~$Q^\circ$.  Using this chart, the cotangent bundle has a natural chart~$Q^\circ\times\mathbb{R}^n$. The complex structure can be chosen so that the points of the polyhedron are the real coordinates of the complex manifold.

Any real analytic function def\/ined on  $Q^\circ\subset\mathbb{R}^n$ has an analytic extension to $Q^\circ_\mathbb{C}\subset\mathbb{C}^n$, where~$Q^\circ_{\mathbb{C}}$ is $Q^\circ_{\mathbb{C}}=Q^\circ\times\mathbb{R}^n$.
In particular, the fundamental solution~(\ref{fundamental}) has an analytic extension on both variables~$x$ and~$x_0$.

Consider the heat equation on the manifold $Q^\circ_\mathbb{C}$
\begin{gather*}
\frac{\partial \nu_{t}^{z_0}(z)}{\partial t}=\frac{1}{2}\Delta \nu_t^{z_0}(z),
\end{gather*}
where $z_0=x_0+\mathrm{i} y_0$ is a point of $Q^\circ_{\mathbb{C}}$ and the Laplacian is taken with respect to the $2n$ real variables $x\in Q^\circ$ and $y\in\mathbb{R}^n$ of $z=x+\mathrm{i} y$. Given the product structure of the chart, the fundamental solution can be calculated as the product of the solution on $Q^\circ$ by the solution on~$\mathbb{R}^n$~\cite{grigoryan06, vanleeuwen09}.

In the Euclidean case, the heat kernel is given by
\begin{gather*}
\nu_t^{z_0}(z)=\frac{1}{(2\pi t)^{n}}\mathrm{e}^{-|z-z_0|^2/2t}.
\end{gather*}
Furthermore, in the case of $Q^\circ_{\mathbb{C}}=Q^\circ\times\mathbb{R}^n$
described above we obtain
\begin{gather}\label{kernelenqc}
\nu_t^{z_0}(z)=\frac{1}{V(2\pi t)^{n/2}} \mathrm{e}^{-|\Im (z-z_0)|^2/2t}\sum_{K\in\mathcal{L}}\mathrm{e}^{\mathrm{i} K\cdot \Re(z-z_0)-K^2t/2}.
\end{gather}
Then we def\/ine the scalar product of holomorphic functions $\phi(z)$ and $\psi(z)$ of $Q^\circ_{\mathbb{C}}$ as
\begin{gather*}
\langle \psi, \phi \rangle_{Q_\mathbb{C}} = \int_{Q_\mathbb{C}}\overline{\psi(z)}\phi(z)\nu_{t/2}^{x_0}(z)\mathrm{d} z.
\end{gather*}
The evaluation in $t/2$ has been done for convenience in order to obtain the isometry~(\ref{isometria}) below.

We call $\mathcal{H}L^2(Q_\mathbb{C},\nu_{t/2}^{x_0})$ to the Hilbert space of square integrable holomorphic functions with this scalar product. In the Euclidean case ($Q=\mathbb{R}^n$) it is the Segal--Bargmann space.

The spaces of holomorphic functions shown in this paper are examples of reproducing kernel Hilbert spaces \cite{hall00a}. In these spaces there is a~function $K(z,\bar{w})$, holomorphic in both arguments, i.e., $K$ is holomorphic in $z$ and antiholomorphic in~$w$ (holomorphic in~$\bar{w}$). For all $\phi\in\mathcal{H}L^2(Q_\mathbb{C},\nu_{t/2}^{x_0})$ the following identity is verif\/ied
\begin{gather}\label{kernelreproductor}
\phi(z) = \int_{Q_\mathbb{C}} K(z,\bar{w})\phi(w)\nu_{t/2}^{x_0}(w) \mathrm{d} w.
\end{gather}
In the Euclidean case the reproducing kernel can be obtained easily \cite{hall00a}
\begin{gather*}
K(z,\bar{w})=\mathrm{e}^{z\bar{w}/t}.
\end{gather*}

We consider linear operators on these spaces. Let $A$ be represented by a kernel $K_A(z,\bar{w})$ as follows
\begin{gather}\label{K_A}
A\phi(z) = \int_{Q_\mathbb{C}} K_A(z,\bar{w}) \phi(w)\nu_{t/2}^{x_0}(w) \mathrm{d} w.
\end{gather}

Given an orthonormal basis of the Hilbert space $\{u_i(\xi)\}$, $i=1,\ldots$, the reproducing kernel can be written
\begin{gather}\label{K}
K(z,\bar{w}) = \sum_{i=1}^\infty u_i(z)\overline{u_i(w)},
\end{gather}
and the corresponding kernel of the operator $A$ is given by
\begin{gather*}
K_A(z,\bar{w}) = \sum_{i=1}^\infty (Au_i(z))\overline{u_i(w)}.
\end{gather*}
Also, the composition of operators is associated with the following kernel
\begin{gather*}
K_{AB}(z,\bar{w}) = \int_{Q_\mathbb{C}} K_A(z,\bar{v}) K_B(v,\bar{w})\nu^{x_0}_{t/2}(v) \mathrm{d} v .
\end{gather*}

\subsection{Isometry between Hilbert spaces}

The spaces $L^2(Q,\rho_t^{x_0})$ and $\mathcal{H}L^2(Q_\mathbb{C},\nu_{t/2}^{x_0})$ are naturally isomorphic. The isomorphism
\begin{gather}\label{isometria}
\mathcal{A}_t\colon \ L^2(Q,\rho_t^{x_0})\to\mathcal{H}L^2(Q_\mathbb{C},\nu_{t/2}^{x_0})
\end{gather}
is given by integration over $Q$ as follows
\begin{gather}\label{transformada}
\mathcal{A}_t f(z)=\int_Q\rho_t^z(x)f(x)\mathrm{d} x,
\end{gather}
where $\rho_t^z(x)$ is the analytic extension of $\rho_t^{x_0}(x)$ in the variable~$x_0$.

The last expression is an isomorphism that can be explicitly tested as follows using the expressions for the kernels of the heat equations
\begin{gather*}
\langle \mathcal{A}_t f, \mathcal{A}_t g \rangle_{Q_\mathbb{C}}  = \int_{Q_\mathbb{C}}\overline{\mathcal{A}_t f(z)}\mathcal{A}_t g(z)\nu_{t/2}^{x_0}(z)\mathrm{d} z \\
\hphantom{\langle \mathcal{A}_t f, \mathcal{A}_t g \rangle_{Q_\mathbb{C}}}{}
 = \int_{Q\times Q}\overline{f(x)}g(x')\int_{Q_\mathbb{C}}\overline{\rho_t^{z}(x)}\rho_t^{z}(x')\nu_{t/2}^{x_0}(z)\mathrm{d} z\mathrm{d} x\mathrm{d} x'\\
\hphantom{\langle \mathcal{A}_t f, \mathcal{A}_t g \rangle_{Q_\mathbb{C}}}{}
= \int_{Q\times Q}\overline{f(x)}g(x') \rho_t^{x_0}(x')\delta(x-x') \mathrm{d} x\mathrm{d} x'
 = \langle f, g \rangle_Q.
\end{gather*}
The integral over $Q_\mathbb{C}$ in the second line is evaluated using the expressions~(\ref{kernelenq}) and~(\ref{kernelenqc}) along with the usual orthogonality relations.

This form of the isometry is the analogous to what Hall calls~$B_t$~\cite{hall94}. There are other forms, however we continue using~(\ref{transformada}) for the purpose of f\/inding the path integral.

\subsection[The $S^1$ case]{The $\boldsymbol{S^1}$ case}

Now we consider the case in which $Q=S^1$. The fundamental solution $\rho_t^{\theta_0}(\theta)$, centered on~$\theta_0$, satisf\/ies the heat equation
\begin{gather*}
\frac{\partial \rho_t^{\theta_0}(\theta)}{\partial t}=\frac{1}{2}\Delta_{S^1}\rho_t^{\theta_0}(\theta)
\end{gather*}
and converges to the Dirac delta $\delta(\theta-\theta_0)$, for $t\rightarrow0^+$.

The function $\rho_t^{\theta_0}$ is given by
\begin{gather*}
\rho_t^{\theta_0}(\theta)=\sum_{k=-\infty}^\infty \rho_k(t) \mathrm{e}^{\mathrm{i} k(\theta-\theta_0)},
\end{gather*}
and therefore the functions $\rho_k$ satisfy
\begin{gather*}
\frac{\mathrm{d}\rho_k(t)}{\mathrm{d}t}=-\frac{1}{2}k^2\rho_k(t),
\end{gather*}
and we f\/inally obtain
\begin{gather}\label{rhos1}
 \rho_t^{\theta_0}(\theta)=\frac{1}{2\pi}\sum_{k\in\mathbb{Z}}\mathrm{e}^{\mathrm{i} k(\theta-\theta_0)-\frac{1}{2}k^2t}.
\end{gather}
Then the scalar product of two functions $f$ and $g$ on $S^1$ is
\begin{gather*}
\langle f,g\rangle = \frac{1}{2\pi}\int_{-\pi}^\pi \sum_{m,n=-\infty}\bar{c}_m d_n \mathrm{e}^{\mathrm{i}\theta(n-m)}\sum_{k=-\infty}^\infty \mathrm{e}^{\mathrm{i} k(\theta-\theta_0)-k^2t/2}\,\mathrm{d}\theta\\
\hphantom{\langle f,g\rangle}{}
= \sum_{m,n=-\infty}^\infty \bar{c}_md_n \mathrm{e}^{-\mathrm{i}(m-n)\theta_0-(m-n)^2t/2}
= \sum_{k=-\infty}^\infty\left(\sum_{j=-\infty}^\infty \bar{c}_j d_{j-k}\right)\mathrm{e}^{-\mathrm{i} k\theta_0-k^2t/2},
\end{gather*}
where $c_j$ and $d_j$ are the Fourier coef\/f\/icients of $f$ and $g$, respectively.

The cotangent space is the cylinder $S^1\times\mathbb{R}$. Thus, the complex manifold $S^1_\mathbb{C}$ can be represented by the chart
$(-\pi,\pi)\times\mathbb{R}$ viewed as the vertical strip in the complex plane.

Holomorphic functions def\/ined on the cylinder are holomorphic functions on $\mathbb{C}$, which are also periodic on the coordinate corresponding to the real part. Using the Cauchy--Riemann conditions it is straightforward to see that they are of the form
\begin{gather*}
\psi(w)=\sum_{l=-\infty}^\infty \psi_l \mathrm{e}^{\mathrm{i} lw}.
\end{gather*}

Given that the heat kernel on a product manifold is the product of the respective heat kernels \cite{grigoryan06, grigoryan09}, in this case in particular the heat kernel on $S^1\times\mathbb{R}$ is obtained from the heat kernel on~$S^1$~(\ref{rhos1}) and the heat kernel on~$\mathbb{R}$~(\ref{rhor}), respectively.

Then, the measure $\nu_t^{z_0}(z)$ in this case is
\begin{gather*}
\nu_t^{z_0}(z)=\frac{1}{2\pi\sqrt{2\pi t}}\mathrm{e}^{-\frac{\Im(z-z_0)^2}{2t}}\sum_{n=-\infty}^\infty \mathrm{e}^{\mathrm{i} n\Re(z-z_0)-n^2t/2}.
\end{gather*}

The Segal--Bargmann transform (\ref{transformada}) of a function $f$ can be obtained easily from his Fourier coef\/f\/icients. It is given by
\begin{gather*}
\psi(z)=\frac{1}{2\pi}\int_{-\pi}^\pi f(\theta)\sum_{n=-\infty}^\infty \mathrm{e}^{\mathrm{i} n(\theta-z)-n^2t/2} \mathrm{d}\theta
 =\sum_{m=-\infty}^\infty c_m \mathrm{e}^{\mathrm{i} mz-m^2t/2}.
\end{gather*}

Given that $\{\mathrm{e}^{\mathrm{i} nx}/\sqrt{2\pi}\}$ with integer $n$ is an orthonormal basis of $L^2(S^1)$, using (\ref{iso}) and (\ref{transformada}) we can obtain $\{\phi_n\}$, which is an orthonormal basis of $\mathcal{H}L^2(S^1,\nu_t^{x_0})$,
\begin{gather*}
\phi_n(z)=\frac{1}{\sqrt{2\pi}}\int_{-\pi}^\pi \frac{\mathrm{e}^{\mathrm{i} nx}\rho_t^z(x)}{\sqrt{\rho_t^{x_0}(x)}}\mathrm{d} x.
\end{gather*}

Then, by equation (\ref{K}) the reproducing kernel is
\begin{gather*}
K(z,\overline{w})=\frac{1}{2\pi}\sum_{k\in \mathbb{Z}}\int_{-\pi}^\pi \int_{-\pi}^\pi\frac{\mathrm{e}^{\mathrm{i} kx}\mathrm{e}^{-\mathrm{i} ky}}{\sqrt{\rho^{x_0}_t(x)\rho^{x_0}_t(y)}}\rho^{z}_t(x)\rho^{\overline{w}}_t(y)\mathrm{d} x\mathrm{d} y
=\frac{1}{2\pi}\int_{-\pi}^\pi\frac{\rho_t^z(x)\rho_t^{\overline{w}}(x)}{\rho_t^{x_0}(x)}\mathrm{d} x.
\end{gather*}

\section{The Feynman path integral}\label{fey}

In this section we propose a def\/inition of the Feynman integral in the holomorphic representation which is suitable for f\/lat manifolds. Here, we take a heuristic motivation for our def\/inition.

First we consider the propagation of the wave function $\phi$ for a small time~$\epsilon$. The propagation is governed by the Schr\"odinger equation with Hamiltonian operator~$H$. We call~$K_H$ to the kernel of this Hamiltonian in the holomorphic representation.

We obtain the evolution operator by exponentiating the Hamiltonian. Applying the evolution operator to~(\ref{kernelreproductor}), using~(\ref{K_A}) and introducing the series expansion of the exponential we obtain
\begin{gather*}
\mathrm{e}^{-\mathrm{i}\epsilon H}\phi(z)  =  \int_{Q_\mathbb{C}}  \left(\sum_{n=0}^\infty\frac{(-\mathrm{i}\epsilon)^n}{n!}K_{H^n}(z,\bar{w})\right) \phi(w)\nu_{t/2}^{x_0}(w)\mathrm{d} w \\
\hphantom{\mathrm{e}^{-\mathrm{i}\epsilon H}\phi(z)}{}
  =  \int_{Q_\mathbb{C}} K(z,\bar{w}) \mathrm{e}^{-\mathrm{i}\epsilon K_H(z,\bar{w})/K(z,\bar{w})} \phi(w)\nu_{t/2}^{x_0}(w)\mathrm{d} w + \epsilon^2\psi(z,\epsilon).
\end{gather*}
Thus, calling $U_{\epsilon}\phi(z)=\mathrm{e}^{-\mathrm{i}\epsilon H}\phi(z)$ and $\tilde{U}_{\epsilon}\phi(z)$ to the last integral, we have
\begin{gather*}
U_{\epsilon}\phi(z)=\tilde{U}_{\epsilon}\phi(z)+\epsilon^2\psi(z,\epsilon).
\end{gather*}

If we divide a time interval $T$ into $n$ equal parts, i.e., $T=n\epsilon$, we have
\begin{gather*}
U_T=\lim_{n\to\infty}U_{T/n}^n,
\end{gather*}
then
\begin{gather*}
U_T\phi(z)=\lim_{n\to\infty}U_{T/n}^n\phi(z)=\lim_{n\to\infty}\left(\tilde{U}_{T/n}^n\phi(z) + \left(\frac{T}{n}\right)^2\Gamma(z,T/n)\right)=\tilde{U}_T\phi(z).
\end{gather*}
We def\/ine a Feynman propagator for a f\/inite time $T$ by dividing the interval $[0,T]$ into $n$ equal subintervals and taking the limit $n\to\infty$. Then, it follows
\begin{gather}\label{propagadorfey}
G(z_T,z_0;T) = \lim_{n\to\infty} \int_{Q_\mathbb{C}^{n-1}}\mathrm{e}^{-\mathrm{i}\frac{T}{n}\sum\limits_{j=1}^n h(z_j,\bar{z}_{j-1})}\prod_{j=1}^n K(z_j,\bar{z}_{j-1})\prod_{j=1}^{n-1}\nu_{t/2}^{x_0}(z_j)\mathrm{d} z_j,
\end{gather}
where we introduce the normal symbol of the Hamiltonian
\begin{gather*}
h(z,\bar{w})=\frac{K_H(z,\bar{w})}{K(z,\bar{w})}.
\end{gather*}
The above expression can also be calculated in the Euclidean case. Here, we take the Hamiltonian
\begin{gather*}
H=z\frac{\partial}{\partial z},
\end{gather*}
corresponding to a renormalized one-dimensional harmonic oscillator. The kernel of this Hamiltonian is $K_H(z,\bar{w})=z\bar{w}\mathrm{e}^{z\bar{w}}$ and the normal symbol $h(z,\bar{w})=z\bar{w}$.

The Feynman propagator (\ref{propagadorfey}) in this case is
\begin{gather*}
G(z_T,z_0;T) = \lim_{n\to\infty}\int_{\mathbb{C}^{n-1}} \mathrm{e}^{\sum\limits_{j=1}^nz_j\bar{z}_{j-1}-\mathrm{i}\frac{T}{n}\sum\limits_{j=1}^nz_j\bar{z}_{j-1}-\sum\limits_{j=1}^{n-1}|z_j|^2}\prod_{j=1}^{n-1}\frac{\mathrm{d} z_j}{2\pi}.
\end{gather*}
Then, by regrouping terms, we have
\begin{gather*}
G(z_T,z_0;T) = \lim_{n\to\infty} \int_{\mathbb{C}^{n-1}} \mathrm{e}^{{z_n\bar{z}_{n-1} - \epsilon\sum\limits_{j=1}^{n-1}z_j\frac{(\bar{z}_{j}-\bar{z}_{j-1})}{\epsilon}-\mathrm{i}\epsilon\sum\limits_{j=1}^{n-1}z_j\bar{z}_{j-1}-\mathrm{i}\epsilon z_n\bar{z}_{n-1}}}\prod_{j=1}^{n-1}\frac{\mathrm{d} z_j}{2\pi},
\end{gather*}
where again $\epsilon=T/n$.

This expression is the known Feynman integral
\begin{gather*}
G(z_T,z_0;T)=\int \mathrm{e}^{z(T)\bar{z}(T)}\mathrm{e}^{\mathrm{i}\int_0^T (\mathrm{i} z(s)\dot{\bar{z}}(s)-h(z(s),\bar{z}(s)))\mathrm{d} s}\mathcal{D}[z(s)],
\end{gather*}
shown in \cite{deligne-et-al99, zinnjustin05}.

\section{Discussion}

The cotangent bundle of a Riemannian manifold admits a complex structure compatible with the symplectic form an the natural lifting of the Riemannian metric if and only if the manifold is f\/lat, i.e., this structure is not integrable unless the curvature is zero, see Appendix~\ref{appendix}.

The restriction on the integrability of the almost-complex structure can be circumvented by changing the metric. In~\cite{gorbunov-et-al05}, in the context of deformation quantization, Gorbunov et al.\ constructed a formally K\"ahler metric on the cotangent space by adding powers of momentum, obtaining an integrable K\"ahler structure.

Furthermore, it has been shown that a complex manifold can be constructed in a neighborhood of the null section of the tangent bundle of a real Riemannian manifold. These are called Grauert tubes \cite{guilleminstenzel91, guilleminstenzel92, lempertszoke91, szoke91}. The complex structures def\/ined therein, called adapted complex structures, are compatible with the symplectic structure leading to a K\"ahler manifold \cite{hallkirwin11}. In certain cases this complex structure exists throughout the tangent, for example when the base manifold is a compact Lie group with a bi-invariant metric.

Our paper is focused on f\/lat Riemannian manifolds, specif\/ically on Euclidean space forms. They present a particular interest in cosmology because they could model the spatial part of f\/lat universe models, see \cite{ellis71, levin02}. The formulas presented in this paper are applicable only to f\/lat manifolds. However the method that leads to the formulation of the path integral is interesting in itself. Moreover, it allows exploring the quantization of space-forms, a topic sparsely discussed in the literature. Finally, the method presented here could be useful for generalizations in future works.

\appendix

\section{Complex structure in phase space}\label{appendix}

It is known that the cotangent bundle of a Riemannian manifold admits a natural complex structure, compatible with the symplectic form and the natural lifting of the Riemannian metric if and only if the manifold is f\/lat~\cite{gorbunov-et-al05, kobayashi69-2}.

First we consider the general case. Let $Q$ be a f\/inite-dimensional orientable real-analytic Riemannian manifold and let $T^*Q$ be the cotangent bundle with projection  $\pi\colon T^*Q\to Q$ and canonical symplectic form $\omega$.

The metric in $Q$ can be naturally lifted to the cotangent space $T^*Q$ as follows. Let $\alpha(t)=(q_1(t),p_1(t))$ and $\beta(t)=(q_2(t),p_2(t))$ be curves on $T^*Q$ such that
\begin{gather*}
\alpha(0)=\beta(0)=(q,p)=m,\qquad \alpha'(0)=V,\qquad\mbox{and}\qquad \beta'(0)=W,
\end{gather*}
i.e., $V$ and $W$ are tangent vectors on $m$. We denote by $\sigma$ the metric on $Q$, and by $\sigma^\sharp$ the isomorphism induced by $\sigma$ between $T^*Q$ and $TQ$. Then we have a metric $G$ on $T^*Q$ given by the following expression
\begin{gather*}
G_m(V,W)=\sigma_q\left(T\pi V,T\pi W\right)+\sigma_q\left(\sigma^\sharp_q\frac{Dp_1}{\mathrm{d} t}(0),\sigma^\sharp_q\frac{Dp_2}{\mathrm{d} t}(0)\right).
\end{gather*}
Here, $T\pi$ is the tangent application of the projection and $D$ represents the covariant derivative.

Thus, we have two dif\/ferent structures on $T^*Q;$ the canonical symplectic structure $\omega$ and the Riemannian metric~$G$.  A third one appears naturally. It is the almost-complex structure~$J$. If $\omega^\sharp\colon T^*T^*Q\to TT^*Q$ is the isomorphism induced by~$\omega$, and $G^\flat\colon TT^*Q\to T^*T^*Q$ the isomorphism induced by~$G$, then~$J$ is given by $J= \omega^\sharp G^\flat$. It means that given the f\/ields~$V$ and~$W$, they verify
\begin{gather*}
G(V,W)=\omega(V,JW).
\end{gather*}
At each point $m$ the map $J_m\colon T_mT^*Q\to T_mT^*Q$ verif\/ies $J_m^2V=-V$. $J$ is compatible with~$\omega$ and~$G$. $(J,G,\omega)$ form what is known as a compatible triple.

We are now interested in f\/inding appropriate local complex coordinates. The complexif\/ied tangent bundle~$T^\mathbb{C}T^*Q$ splits as follows
\begin{gather*}
T^\mathbb{C}T^*Q=T^{(1,0)}T^*Q\oplus T^{(0,1)}T^*Q.
\end{gather*}
Here $T^{(1,0)}T^*Q$ and $T^{(0,1)}T^*Q$ are the images of the projections $\Pi^+$ and $\Pi^-$ given by
\begin{gather*}
\Pi^\pm = \frac{1 \mp \mathrm{i} J}{2}.
\end{gather*}

If we choose a vector $V=(\dot{q}^1,\ldots,\dot{q}^n,\dot{p}_1,\ldots,\dot{p}_n)\in T_mT^*Q$, then $\Pi^+$ is a natural isomorphism between the real tangent space $T_mT^*Q$ and the {\it holomorphic tangent space} $T^{(1,0)}_mT^*Q$. Explicitly, we have
\begin{gather*}
\Pi^+V=\dot{z}^i\frac{\partial}{\partial z^i},
\end{gather*}
where $\dot{z}^i=\dot{q}^i+\mathrm{i}\sigma^{im}(\dot{p}_m-p_k\Gamma_{ml}^k\dot{q}^l)$.

The corresponding holomorphic vector f\/ields are
\begin{gather*}
\frac{\partial}{\partial z^i} = \frac{1}{2}\left(\frac{\partial}{\partial q^i} + p_k\Gamma^k_{ij}\frac{\partial}{\partial p^j} - \mathrm{i}\sigma_{ij}\frac{\partial}{\partial p^j}\right),
\end{gather*}
where $\sigma_{ij}$ and $\Gamma^k_{ij}$ are the matrix coef\/f\/icients of the metric and the Christof\/fel symbols respectively. Henceforth, the Einstein summation convention is used.

If we take the Lie bracket of the above-mentioned f\/ields we obtain
\begin{gather*}
\left[\frac{\partial}{\partial z^i},\frac{\partial}{\partial z^j}\right] = \mathrm{i} R^m_{kij}p_m\sigma^{lk}\left(\frac{\partial}{\partial z^l}-\frac{\partial}{\partial\bar{z}^l}\right).
\end{gather*}
Then, by the Niremberg--Newlander theorem \cite{kobayashi69-2}, the distribution is integrable if and only if the curvature tensor of the metric~$\sigma$ is identically null, see Gor\-bu\-nov et al.~\cite{gorbunov-et-al05}.

\subsection*{Acknowledgements}

We thank Hern\'an Cendra for his reading of the manuscript and useful
suggestions. This work was supported by the Universidad Nacional del Sur (Grants PGI 24/L085, PGI 24/L086 and PGI 24/ZL10).

\pdfbookmark[1]{References}{ref}
\LastPageEnding

\end{document}